\documentclass[manuscript]{aastex}

\slugcomment{070516/070820}

 \shorttitle{Does    Sub-millisecond  Pulsar
 XTE J1739-285 }
 \shortauthors{C. M.  ZHANG  et al.}

\begin{document}

\def\gapprox{\lower.4ex\hbox{$\;\buildrel >\over{\scriptstyle\sim}\;$}}
\def\lapprox{\lower.4ex\hbox{$\;\buildrel <\over{\scriptstyle\sim}\;$}}
\def\be{\begin{equation}}
\def\ee{\end{equation}}
\def\bea{\begin{eqnarray}}
\def\eea{\end{eqnarray}}

\def\lesssim{\mathrel{\hbox{\rlap{\hbox{\lower4pt\hbox{$\sim$}}}\hbox{$<$}}}}
\def\gtrless{\mathrel{\hbox{\rlap{\hbox{\lower3pt\hbox{$<$}}}\hbox{$>$}}}}

\def\cenx4{{Cen~X$-$4}}
\def\aql{{Aql~X$-$1}}
\def\1e{{1E 1207.4-5209}}
\def\exo{{EXO 0748-676}}
\def\saxj{{SAX J1808.4$-$3658}}
\newcommand{\ud}[2]{\mbox{$^{+ #1}_{- #2}$}}

\def\c{\cite}

\def\et{ {\it et al.}}
\def\lan{ \langle}
\def\ran{ \rangle}
\def\ov{ \over}
\def\ep{ \epsilon}

\def\et{ {\it et al.}}
\def\la{ \langle}
\def\ra{ \rangle}
\def\ov{ \over}
\def\ep{ \epsilon}

\def\mdot{\ifmmode \dot M \else $\dot M$\fi}    
\def\mxd{\ifmmode \dot {M}_{x} \else $\dot {M}_{x}$\fi}
\def\med{\ifmmode \dot {M}_{Edd} \else $\dot {M}_{Edd}$\fi}
\def\bff{\ifmmode B_{f} \else $B_{f}$\fi}

\def\apj{\ifmmode ApJ\else ApJ\fi}    
\def\apjl{\ifmmode  ApJ\else ApJ\fi}    %
\def\aap{\ifmmode A\&A\else A\&A\fi}    %
\def\mnras{\ifmmode MNRAS\else MNRAS\fi}    %
\def\nat{\ifmmode Nature\else Nature\fi}
\def\prl{\ifmmode Phys. Rev. Lett. \else Phys. Rev. Lett.\fi}
\def\prd{\ifmmode Phys. Rev. D. \else Phys. Rev. D.\fi}

\def\ms{\ifmmode {\rm M_{\odot}} \else ${\rm M_{\odot}}$\fi}    
\def\na{\ifmmode \nu_{A} \else $\nu_{A}$\fi}    
\def\nk{\ifmmode \nu_{K} \else $\nu_{K}$\fi}    
\def\ns{\ifmmode \nu_{{\rm s}} \else $\nu_{{\rm s}}$\fi}
\def\no{\ifmmode \nu_{1} \else $\nu_{1}$\fi}    
\def\nt{\ifmmode \nu_{2} \else $\nu_{2}$\fi}    
\def\ntk{\ifmmode \nu_{2k} \else $\nu_{2k}$\fi}    
\def\dnmax{\ifmmode \Delta \nu_{max} \else $\Delta \nu_{2max}$\fi}
\def\ntmax{\ifmmode \nu_{2max} \else $\nu_{2max}$\fi}    
\def\nomax{\ifmmode \nu_{1max} \else $\nu_{1max}$\fi}    
\def\nn{\ifmmode \nu_{\rm NBO} \else $\nu_{\rm NBO}$\fi}    
\def\nh{\ifmmode \nu_{\rm HBO} \else $\nu_{\rm HBO}$\fi}    
\def\nqpo{\ifmmode \nu_{QPO} \else $\nu_{QPO}$\fi}    
\def\nz{\ifmmode \nu_{o} \else $\nu_{o}$\fi}    
\def\nht{\ifmmode \nu_{H2} \else $\nu_{H2}$\fi}    
\def\ns{\ifmmode \nu_{s} \else $\nu_{s}$\fi}    
\def\nb{\ifmmode \nu_{{\rm burst}} \else $\nu_{{\rm burst}}$\fi}
\def\nkm{\ifmmode \nu_{km} \else $\nu_{km}$\fi}    
\def\ka{\ifmmode \kappa \else \kappa\fi}    
\def\dn{\ifmmode \Delta\nu \else \Delta\nu\fi}    

\def\vk{\ifmmode v_{k} \else $v_{k}$\fi}    
\def\va{\ifmmode v_{A} \else $v_{A}$\fi}    %
\def\vf{\ifmmode v_{ff} \else $v_{ff}$\fi}    

\def\rs{\ifmmode {R_{s}} \else $R_{s}$\fi}    
\def\ra{\ifmmode R_{A} \else $R_{A}$\fi}    
\def\rso{\ifmmode R_{S1} \else $R_{S1}$\fi}    
\def\rst{\ifmmode R_{S2} \else $R_{S2}$\fi}    
\def\rmm{\ifmmode R_{M} \else $R_{M}$\fi}    
\def\rco{\ifmmode R_{co} \else $R_{co}$\fi}    
\def\ris{\ifmmode {R}_{{\rm ISCO}} \else $ {\rm R}_{{\rm ISCO}} $\fi}
\def\rsix{\ifmmode {R_{6}} \else $R_{6}$\fi}
\def\rinfty{\ifmmode {R_{\infty}} \else $R_{\infty}$\fi}
\def\rinfsix{\ifmmode {R_{\infty6}} \else $R_{\infty6}$\fi}

\def\rxj{\ifmmode {RX J1856.5-3754} \else RX J1856.5-3754 \fi}
\def\1739{\ifmmode {XTE  J1739-285} \else XTE  J1739-285 \fi}
 \def\exo{\ifmmode {EXO 0748-676} \else EXO 0748-676 \fi}
\def\sax{\ifmmode {SAX J1808.4-3658} \else SAX J1808.4-3658 \fi}

\title{ \bf Does    Sub-millisecond  Pulsar
 XTE J1739-285  Contain a  Low Magnetic  Neutron Star or Quark  Star ? }


\author{C. M. Zhang$^{1}$, H.X. Yin$^{1}$,  Y.H.
Zhao$^{1}$,  Y.C. Wei$^{1}$, X.D. Li$^{2}$ }

\affil{1. National Astronomical Observatories,
 Chinese Academy of Sciences, Beijing 100012, China, zhangcm@bao.ac.cn\\
 2. Department of Astronomy,
 Nanjing University, Nanjing 210093, China,
 lixd@nju.edu.cn }


\begin{abstract}
 With the  possible detection  of the fastest spinning nuclear-powered
 pulsar XTE  J1739-285 of frequency 1122 Hz (0.8913 ms), it arouses
 us to constrain the mass   and radius   of  its  central compact object
 and to imply  the stellar  matter compositions: neutrons or quarks.
 Spun-up by the accreting   materials  to such a high rotating speed,
  the compact star should have either  a   small radius or short
 innermost stable circular orbit. By the empirical relation between
  the upper  kHz quasi-periodic oscillation  frequency and  star
  spin frequency,  a strong constraint on mass and radius is obtained  as
 1.51 solar masses  and 10.9 km, which  excludes most equations
 of states  (EOSs) of normal neutrons and strongly hints the star
 promisingly to be a strange  quark star.
 Furthermore, the star magnetic  field  is  estimated to be about
 $4\times10^{7} (G) < B < 10^{9}  (G) $,   which  reconciles  with
  those of millisecond  radio   pulsars,  revealing the clues of
  the evolution linkage of two types of  astrophysical objects.
  \end{abstract}

\maketitle


\section{Introduction}

The most rapidly  spinning astrophysical object in the universe,
named XTE J1739-285, rotating  1122 times per second,  has been
declaimed to be detected  recently in  the accreting X-ray binary
system (Kaaret et al. 2007), which is the first sub-millisecond
pulsar (spin period 0.89 ms) coming into our view  since the
discovery of the first radio pulsar by Jocelyn Bell and Anthony
Hewish  forty years ago (Hewish, Bell \& Pilkingston et al. 1968) if
the 1122 Hz is a spin frequency. 

  However it is stressed that the  1122 Hz burst oscillation
frequency is detected in the only one burst but not the whole set of
bursts (Kaaret et al. 2007),   so  this frequency still needs the
further confirmation in the future burst detections. At the present
time we take this 1122 Hz frequency as a tentative case,  and based
on this
  our  investigations of its applications  are processed.
  If the declared spin frequency 1122 Hz is fully confirmed,  then
  this  compact object breaks through the record
 of the fastest  radio pulsar
with the spinning frequency of   716 Hz (1.39 ms)    discovered
recently
 (Hessels et al. 2006).  Strikingly, such a high spinning makes
 the edge  velocity at the  radius of  42.6 km  the speed of light,
 which quantitatively constraints on the  star size and its emission
 region.   If  this declaimed  spin detection  is true,  it is  interesting
  that such a high spinning does not break up the star,
  which provides us a unique  opportunity  to explore the
matter compositions of the central object,  neutrons or their
constituent quarks  that have been never seen as free particles to
date.
 No doubt,  such a high  rotating frequency  1122 Hz should
influence  on the EOS of compact object and  its
mass-radius relation (Lavagetto et al. 2006; Bejger et al. 2006).
In fact, the importance of detecting a sub-millisecond rotating pulsar as
a way of discriminating  the star nuclear compositions has been
proposed by several authors
(Phinney \& Kulkarni 1994; Burderi \& D¡¯Amico 1997; Burderi et al. 1999, 2001).
 The formation of a very fast spinning pulsar, however, depends sensitively on
 the history of the compact object in the binary and on the evolution of its
  magnetic field  (Possenti et al. 1998).

 The source  XTE J1739-285, exhibiting  as a transient X-ray burst source, firstly
 discovered in 1999 by the satellite {\it Rossi X-ray Timing Explorer (RXTE)},
 is recently reported  that the burst oscillation frequency is detected
 at 1122 Hz (Kaaret et al. 2007), which is identified as a spin frequency
  because in the accretion-powered pulsar SAX J1808.4-3658
 both  a pulsation frequency and a burst frequency are  detected at 401 Hz
 (Chakrabarty et al. 2003; van der Klis 2006; Yin, Zhang \& Zhao, et al. 2007).
The central object of \1739 lies in a low mass X-ray binary (LMXB) and
is believed to be a dense star with the incredible density,
 like what a solar mass is trapped into a area of 10 km as a result of the
 supernova explosion of some massive star.
 The  magnetic field of this type of object is about
  $\sim10^{8-9}$ Gauss, which experiences
   deduction  from the original value  $\sim10^{12}$ Gauss
 by sucking the accreting materials from the disk fed by its companion
 (e.g. Bhattacharya \& van den Heuvel 1991).
 The rotating of such a  magnetic object will be responsible for its pulsation.

Measured  mass and radius of compact object  can be used to study
the strong nuclear force of   matter  or the relation between  the
matter pressure and density, i.e.  equation of state (EOS) (Lattimer
\& Prakash  2001, 2004, 2006; Glendenning 2000).
 Understanding the behavior of matter at the super-high
density has a special priority in astrophysics because such an
extreme condition cannot be imagined in the laboratory on   Earth.
 The soft (stiff) EOSs predict lower (higher) pressures for
  same  density,  and most of these EOSs give the upper limit
  of star mass to be about 1.6 \ms.  If  a more massive mass is inferred,
  for instance  the star mass M $\sim$ 2.1 \ms ~in \exo, then the soft EOS
  is ruled out (\"Ozel 2006). {\bf Three decades ago, the possibility  of
  quark star was   proposed   and  studied by Itoh (1970),
  Chapline \&  Nauenberg (1977)  and Witten (1984).}
  Later on, the masses   of quark stars have been   calculated
    to primarily reconcile
  the  mass ranges of soft EOSs of NSs,  but combined with
  the smaller radii (e.g. Lattimer \& Prakash  2001, 2004, 2006;
   Dey et al. 1998;  Haensel et al. 1986; Alcock et al. 1986).
  Combined with star mass-radius relation and other  properties,
   the possible candidates of quark  stars have  been  investigated,
   such as the X-ray source \rxj (Drake et al. 2002), accretion-powered
    X-ray pulsar \sax (Li et al. 1999), the X-ray burst
    sources  GRO J1744-28 (Cheng et al. 1998) and  4U 1820-30 (Bombaci 1997),
     as well as the $\gamma$-ray bursters (Haensel et al. 1991).
    Moreover, the   frequency distribution of X-ray neutron stars may
    reveal that a quark phase transition
    resulting from the changing central density induced by the changing
    spin (Glendenning \& Weber 2001). From the inferred small `apparent radius',
    one concludes the star inside 1E 1207.4-5209 may be composed of the
     strange quark matters (e.g. Xu 2005);   the emission properties of
     strange quark star has been explored recently by Melrose et al.
     (2006).
However,  as pointed out,  the astrophysical
 observations   could not  unambiguously distinguish the  quark
stars from normal neutron stars (Lattimer \& Prakash 2004, 2006;
  Page \& Reddy 2006;  Watts \&  Reddy 2007).

The neutron star (NS)  mass  can be measured with  high accuracy in
the binary radio pulsar system (e.g. Kaspi,  Taylor  \& Ryba 1994),
 for instance, the masses  of
  double pulsar  PSR J0737-3039 M=1.337(5) $\ms$ and M=1.250(5) $\ms$
  (Lyne et al. 2004).
  Until now, around fifty NS masses have been measured
  with the averaged value of $\sim$1.4 \ms (Lattimer \& Prakash  2004, 2006),
  from the possible minimum mass 0.97$\pm$0.24 \ms
   (Jonker et al. 2003) to maximum value 2.1$\pm$0.2 \ms
  (Nice et al. 2005).  Unlike the situation of NS mass measure, there is no
effective way of acquiring  an accurate NS radius directly, thus the
star  EOS can be only estimated by some measures  with errors (Zhang et al. 2007).
 Then  some $M-R$ relations can be  measured to derive $M$ and $R$
constraints (\"Ozel 2006; Miller 2002), for instance, the ``apparent
radius" estimated from the thermal emission of perfect blackbody
 (Truemper et al. 2004;  Burwitz et al. 2003; Rutledge et al. 2001;
 Burwitz et al. 2001; Haensel 2001),
the gravitational redshift of the spectral lines  (Cottam et al. 2004), and
 the star magnetosphere limit of \saxj  (Burderi \& King 1998),
etc.
As declaimed, it is important to confine the challenging attempts to measure
 the NS radius to the most reliable data and methods,
 otherwise we will continue to produce  the radius values with the uncertainty
   of a factor of two or so, which is not enough to constrain EOS of NS matter
   (Truemper 2007).

\section{ Mass and radius estimation of \1739}

\subsection{  Constraints on mass and radius by the spin frequency}

Located about 35,000 light-years away from Earth, the burst source
\1739 lies in   an accreting binary system,  where the orbital
materials in the star inner disk of  radius $r$  process the circle
motion
   with Keplerian
frequency $\nk$ (van der Klis 2006; Stella \& Vietri 1999; Miller et
al. 1998), \be \nk = \sqrt{{GM\over 4\pi^{2}r^{3}}} = 1833~ {\rm
(Hz)} ({M\ov\ms})^{1/2}({R\ov10{\rm km}})^{-3/2}\, ,\label{nt} \ee
which should be bigger than the spin frequency $\ns$   because the
star in LMXB is experiencing  the spin-up phase (e.g. van der Klis
2006) .

\begin{figure}
\begin{center}
\includegraphics[width=12cm]{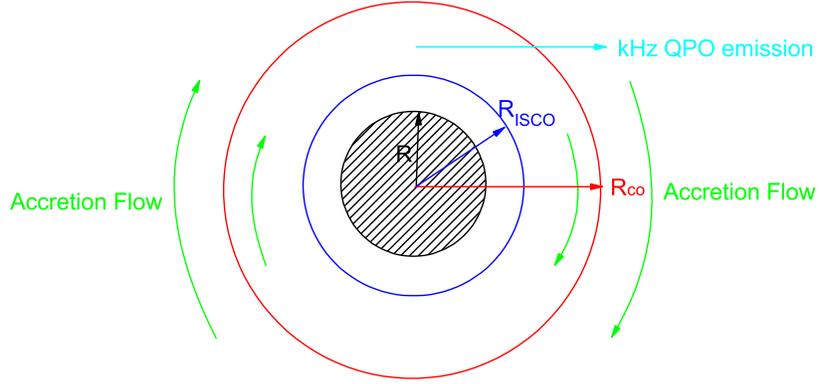}
\end{center}
 \caption
 {  Illustrative diagram  of ISCO and corotation radius,
  where the  star surface  inside the  ISCO is assumed.
 The ISCO radius $\ris$ is three  Schwarzschild radii ( $\ris=6GM$), and the corotation
 radius $\rco$
 is determined by the condition that the Keplerian frequency equals
  spin frequency at $\rco$ as described in Eq. (\ref{rco}), i.e.
  $\nk(\rco)=\ns$.  ISCO is the innermost stable
  circular orbit, so the disk Keplerian flow should occur outside ISCO, implying
  the kHz QPO frequency and spin frequency to be less than the frequency at ISCO.
  The compact star in LMXB is experiencing the spin-up by the accreting  matter,
  so the magnetosphere-disk boundary  should be inside the corotation radius but
  outside ISCO.} \label{isco-rco}
\end{figure}

\begin{figure}
\begin{center}
\includegraphics[width=12cm]{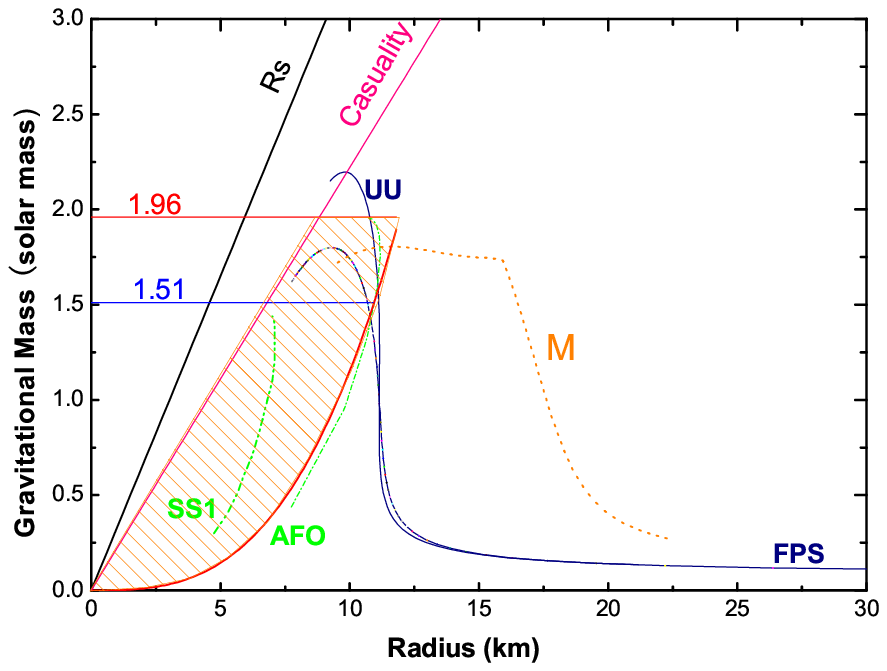}
\end{center}
 \caption
 { Mass-radius
diagram of \1739. Five representative EOSs are shown (for their
definitions, see Lattimer \& Prakash 2001; Cook et al. 1994): stars
containing
  quark matter (green color curves labeled SS1 and AFO), stars
made of normal neutron matter (navy color curves labeled UU and
FPS), and star with the pion  condensation  transition (orange color
curve labeled M). The straight lines represent the constraints by
general relativity and causality conditions,  i.e.  $R=2GM$ and
$R=2.9GM$, respectively (Lattimer \& Prakash 2004; 2006); the parabola
curve represents the rotational constraint, as described in
Eq.(\ref{ns}). The horizontal lines M=1.96 $\ms~$  and M=1.51 $\ms~$
are explained in the text.  The shadowed area
 stands for the possible $M-R$ range  of \1739, which covers
 the EOSs of  the  quark matter and normal neutron
 ($M<1.96 \ms$), or EOS of the quark matter  ($M<1.51 \ms$). } \label{mr}
\end{figure}

 The Keplerian motion will end at  the innermost stable
 circular orbit (ISCO) if the star surface is inside ISCO,
  so the spin-up process will be invalid
there.  The formation of the spin frequency of star in LMXB is due
to the spin up of the accreted matter in the Keplerian motion, thus
the Keplerian orbital frequency in the inner magnetosphere-disk
boundary must exceed over the spin frequency (e.g. Bhattacharya \&
van den Heuvel 1991; Cheng \& Zhang 2000; Lamb \& Boutloukos 2007).
Therefore,  it is usually believed that the maximum Keplerian
frequency occurs at ISCO with the radius ${\ris=6GM}$
 (e.g. Miller et al. 1998; Zhang et al. 1998; see the illustration diagram
 Fig. \ref{isco-rco}), or ISCO lies inside    the corotation radius
$\rco$ at where the Keplerian frequency equals the star  rotation
spin frequency, i.e. $\ris\leq\rco$,
 \be{\rco\ov10{\rm km}}=({M\ov\ms})^{1/3}({1833\ov\ns})^{2/3}\;.
\label{rco} \ee Thus one has  a mass constraint in the following (Miller et al. 1998)
 \be {M\ov\ms} \leq {2200 ({\rm Hz})\ov\ns}\;. \label{mass} \ee
As declaimed, for the given star mass and radius, the permitted
maximum spin frequency is   as follows (Lattimer \& Prakash 2004, 2006)
 \be \ns \leq  1045~ {\rm (Hz)}
({M\ov\ms})^{1/2}({R\ov10{\rm km}})^{-3/2}\;. \label{ns} \ee
Therefore, from Eq.(\ref{mass}) and Eq.(\ref{ns})  the mass-radius
constraints can be obtained,  which are plotted in Fig.~\ref{mr}.

For comparisons, several EOS curves  are plotted in Fig.~\ref{mr} as
well.
  In the shadowed area of Fig.~\ref{mr} with  $M<1.96 \ms$ and $R<11.9\; km$,
    where the meanings of
 its boundaries are indicated in the figure caption, we find that the
 too stiff EOSs  are strongly  excluded.

 \subsection{  More stringent constraints on M and R by the kHz QPOs}

 However, if the higher twin kHz QPOs  are detected in \1739, as
derived from the other twin kHz QPO sources,  we could even give a
stronger constraint on the star mass upper limit.
 In theory, the  Keplerian frequency of   orbital matter, to which
  the upper kHz QPO frequency is identified,
  should be greater than the spin  frequency because of the
   spinning up of star by  accretion
  (Shapiro \& Teukolsky 1983).  From RXTE detections, the twin kHz QPOs,
  the upper and lower frequencies, are often found in the Fourier power spectra
  (van der Klis 2006),  and it is also noticed  that,
   for the  known nuclear-powered millisecond
 pulsars with the simultaneously detected twin kHz QPO frequencies
 shown in Table 1,  their  upper kHz QPO frequencies,
 usually identified as the Keplerian orbital frequency
 (van der Klis 2006; Stella \& Vietri 1999;
  Miller et al. 1998),  are all at least  1.3
 times bigger than their spin frequencies (see Table 1;
 Yin, Zhang \& Zhao et al. 2007),   which implies the inner  orbital
  accreting matter to enter into
   the region enclosed by the corotation radius.  If this empirical
    relation is also valid for  \1739  though no  twin
kHz QPOs have been simultaneously detected in this source (Kaaret et
al. 2007),   one can conclude that the minimum upper frequency of
twin kHz QPOs  of \1739 is bigger than about 1459 (=1.3$\times$1122)
Hz. Henceforth,  the  upper limit of mass constraint  by
Eq.(\ref{mass})  is changed to be  M=1.51\ms.
 Interestingly,  in  the shadowed area of Fig.~\ref{mr} with $M<1.51 \ms$ and $R<10.9\; km$,
 almost all plotted EOSs of  normal neutrons and  pion  condensation
transition   are
 excluded, and only the model  SS1  for quark matter (Dey et al. 1998) is possible.
 Accordingly,  the compact object in \1739 is more
 probably composed of  quark matter, if the declaimed higher kHz QPO frequency is
 detected.
 Perspectively,  these results  sheds light on the study of sub-nuclear physics. In
addition, it would be meaningful to detect the higher kHz QPOs than
1500 Hz in \1739, revealing a smaller mass value than 1.47 \ms
inferred from Eq.(\ref{mass}), which will indispensably exclude
EOSs of normal neutron matters. Furthermore, the $M-R$ constraints
would be considerably improved if the star mass  or some
 $M-R$  relations, e.g. the gravitational redshift and
 thermal emission,  were known.
  In any case, \1739 is a dramatically particular  probe of
exotic matter in extreme environments that cannot be achieved on
Earth. While, as a conclusion, the sub-millisecond pulsar will
definitely exclude  the too stiff EOSs, implying that the star
matter is highly incompressible
 or the big stars cannot endure too fast spin frequencies without
breaking apart.

\section{  On the magnetic field constraints  of \1739} The   magnetic
field of NS in the binary system
 can be measured directly by the cyclotron lines in the magnitude
 order of $\sim 10^{12}$ Gauss (Truemper et al. 1978), however the radio
 pulsars and low magnetic NSs in low mass X-ray binaries (LMXBs)
  have no such priorities.
The magnetic field strength of radio pulsar is estimated by the
assumption of magnetic  dipole radiation accounting for its spin
down, so the approximation is aroused by  exploiting  the presumed
star parameters,  i.e. M=1.4 \ms ~ and R=15 km
(e.g. Manchester \& Taylor 1977). As for
the estimation of star
  magnetic field strength $B$  in LMXB,  it
is derived  by the definition of  magnetosphere radius, which is
also inaccurate   because of the unknown star  mass and radius
(e.g. Shapiro \& Teukolsky 1983; Burderi et al. 1996).
While, the situation will be improved if
one can infer the star  mass and radius of \1739.
 In  the accreting binary system, the magnetosphere radius
$\rmm$ is about an half Alfv\'en radius and
can be conveniently written as (Shapiro \& Teukolsky 1983),
 $\rmm = 0.43\times 10^{6} (cm) ({B\ov10^{8}G})^{4/7}
 ({\mdot\ov10^{18} g/s})^{-2/7}({M\ov\ms})^{-1/7}({R\ov10{\rm
 km}})^{12/7}$, or equivalently 
 \be B = ({\rmm\over R})^{7/4}\bff\;,  \ee
 \be \bff = 4.4 \times10^{8}\; (G)\;
  ({\mdot\ov10^{18} g/s})^{1/2}  ({M\ov\ms})^{1/4} ({R\ov10{\rm
 km}})^{-5/4}\;, \ee
 where $\mdot$ is the accretion rate and $\bff$ is a field strength
  when $\rmm=R$ ($\bff$ is named as a bottom field, see, e.g.
   Zhang \& Kojima 2006; Burderi et al. 1996).
 With the inequality $R<\rmm<\rco$, we obtain the magnetic field
 constraint as follows,
 \be
\bff<B<({\rco\over R})^{7/4}\bff\;.
 \ee
 For \1739, by means of the approximated conditions $2.9GM<R<11.9 {\;} km$,
$M<1.96 \ms$ and Eq.(\ref{ns}) where the radius 2.9 GM is given by the casuality
condition of general relativity (Lattimer \& Prakash 2004, 2006),
 we obtain  the lower and  upper limits of magnetic field of \1739
as, \be B > 4.2 \times10^{8} (G) ({\mdot\ov10^{18} g/s})^{1/2}\;,
\ee and \be  B < 9.97
\times10^{9}(G)({M\ov\ms})^{-13/6}({\mdot\ov10^{18}g/s})^{1/2}\;.
\ee
 \1739 is identified as a less luminous Atoll source (Kaaret et al. 2007),
 so its long-term accretion rate should be as low as
 those of the usual Atoll sources, i.e.  $\mdot\sim10^{16} g/s$
 (on the definition of Atoll source, see Hasinger \& van der Klis 1986).
 In addition, if the star  mass of \1739 is set to be about
$\sim 1 \ms$,  then its  lower and upper limits of magnetic field
strength are approximately given, $4 \times10^{7} (G) < B <
10^{9}(G)$,
 which is   compatible with   the approximately  derived field
  strength $\sim10^{8}$ Gauss
  of millisecond radio pulsar (Bhattachaya \& van den Heuvel 1991;
  van den Heuvel \& Bitzaraki 1995).
  Therefore, the similar magnetic fields and spin frequencies of both
  the compact objects in LMXBs  and millisecond pulsars  in the binary
  systems strongly hint their
  relevant evolutionary linkage (van den Heuvel 2004).   However, the
  magnetic field as low as  $\sim 4\times10^{7}$ Gauss has not yet
been discovered.
   In the accreting binary system, it has long been  believed that
    NS spin is accelerated  by the
 accretion (Alpar et al. 1982; Radhakrishnan, \& Srinivasan 1982)
and its field decays (Bhattachaya \& van den Heuvel 1991;
 van den Heuvel \& Bitzaraki 1995).  The spin and magnetic
field of \1739 strongly reveals the evidence that the compact object
in LMXB is the progenitor of millisecond radio pulsar, which
  strengthens our understanding on  the formation and evolution of such a
striking spinning  object. On the other hand,  the  evolutionary
linkage also means that
 some millisecond pulsars are involved in the quark matters inside stars.
 Apparently,    to confirm  the spin frequency and detect the higher
 kHz QPO frequency of \1739  have the  preferential importance  to
 set the stringent limit   on the physical parameter  of its compact object.


\begin{table}
\begin{minipage}{\linewidth}
{Table 1. List of the low-mass X-ray binaries with
 the simultaneously detected twin kHz QPO and spin frequencies.}
\begin{center} \begin{tabular}{lcccc}
\hline \hline Sources$^{(1)}$  &
$\nt$(Hz)$^{(2)}$ & $\nu$$_{s}$(Hz)$^{(3)}$ &
 $\nu_{2min}/\nu_{s}$$^{(4)}$ & Refs. \\
\hline
4U 1608-52            & 802-1099   &  619   & 1.3 &[1]\\
4U 1636-53            &  971-1192 &  581   & 1.7 &[2]\\
4U 1702-43             & 1055   &  330   & 3.2&[3]\\
4U 1728-34           & 582-1183   &  363   & 1.6&[4]\\
KS 1731-260           & 1169   &  524   & 2.2&[5]\\
4U 1915-05            & 514-1055 &  270   & 1.9&[6]\\
XTE J1807-294       & 353-587  &  191  & 1.8&[7]\\
SAX J1808.4-3658         &  694   &  401  & 1.7&[8]\\
\hline \hline
\end{tabular}
\end{center}
\vskip 0.1cm
\begin{tabular} {p{\linewidth}}{
  $^{(1)}$: On the QPO data, see Belloni et al. 2005,
  van der Klis (2006) and Zhang et al. 2006,  the original references therein;
  $^{(2)}$: upper  kHz QPO frequency;
  $^{(3)}$:   On the spin frequency, see also Chakrabarty 2004; Strohmayer \& Bildsten 2006;
  van der Klis 2006;
            Lamb \& Boutloukos 2007; Yin, Zhang \& Zhao,  et al. 2007,
             the original references therein;
  $^{(4)}$: the ratio between the minimum upper kHz QPO frequency and spin frequency.
   [1]: Hartman et al. 2003; [2]: Wijnands et al.
  1997; [3]: Markwardt et al. 1999; {4}: Strohmayer et al. 1996; [5]: Smith et al. 1997;
  [6]: Galloway et al. 2001; [7]: Markwardt et al. 2003; [8]: Wijnands \& van der Klis
  1998.}
\end{tabular}
\end{minipage}
\end{table}

\section{Consequences and Discussions}

In the paper, we  take the burst oscillation frequency of 1122 Hz of
\1739 to be  the NS spin rate, however it still needs the further
confirmation (e.g. Galloway 2007).  At first, it is detected in the
brightest burst
 with the significant at the 99.96\% confidence level, not in
a whole set of six bursts (Kaaret et al. 2007). Secondly, from a
statistical point of view, the frequency of 1122 Hz is too far apart
from the arrange of the previously known spin frequencies of LMXBs,
from the minimum value of 45 Hz (Villarreal \& Strohmayer 2004) to
maximum 619 Hz (Hartman et al. 2003), centered at $\sim$ 400 Hz
(Yin, Zhang \& Zhao et al. 2007), and it is much higher than the
fastest spin frequency of radio pulsar 716 Hz (Hessels, Ransom \&
Stairs, et al. 2006). Therefore, we  take this spin frequency of
1122 Hz as a tentative detection, or it might be considered that the
signal of 1122 Hz  is merely a candidate burst oscillation (e.g.
Galloway 2007). In addition, we emphasize that this spin  frequency
is not yet verified totally  and the present observation is not
statistically as sound as for those of other LMXB's (see, e.g.,
Strohmayer  \& Markwardt 2002 for 4U 1636- 536; Strohmayer,
Markwardt  \& Swank et al. 2003 for XTE J1814-338).

 By means of this assumed highest spin frequency,
 we can constrain  on the NS mass and radius, which indicates
 the frequency of 1122 Hz to be close to the centrifugal breakup limit
for some equations of state of nuclear matter (e.g. Burgio, Schulze
\& Weber 2003). Furthermore, if the empirical relation on the kHz
QPO frequency and spin frequency is presumed to be possible (see
Table 1), then we can infer the higher kHz QPO frequency than 1500
Hz, which will exclude the EOSs of normal neutrons and just leave
those of the strange quark matters possible. Therefore, the
motivations for the physical
 properties of the super dense nuclear matters  are enhanced (e.g.
  Page \& Reddy 2006;  Watts \&  Reddy 2007).

With the known spin frequency, combined with  the constrained NS
mass and radius, the corotation radius of NS in \1739 is inferred,
by which the constraints of NS  magnetic field strength is derived
with the observationally estimated  mass accretion rate. The similar
estimation of NS magnetic field has been obtained for \sax (Wijnands
\& van der Klis 1998), where the NS mass and radius are assumed,
however
 in this paper we exploit  the constrained mass and radius of NS
 to derive the ranges of NS magnetic field of \1739, which presents
 the upper and lower limits of magnetic field.

\vskip 0.3cm
 Acknowledgements:
 We are grateful for  the  helpful discussions
 with S. Boutloukos, V. Burwitz, K.S. Cheng, G. Hasinger,
  J. Horvath, N. Itoh, B. Kiziltan, D. Menezes,  J. Truemper, S. Reddy, J. Bell, E.P.J. van
  den Heuvel, N. Rea, F. Ozel, P.C.C. Freire,  R.
  Rutledge,   D. Lai,  T. Lu,  Q.H. Peng, G.J. Qiao,
  X.J. Wu,  R.X. Xu, and Z.R. Wang.
We also thank D. Galloway,  M. Mendez, T. Strohmayer, and P. Kaaret
for the discussion of the detection of spin  frequency 1122 Hz of
\1739.
 This research has been  supported by the  innovative project
 of CAS  and NSF of   China.
  We are very grateful for the critic comments from the anonymous
  referee.
%


\end{document}